\documentclass[prl,showpacs,twocolumn]{revtex4}
\usepackage{times}
\usepackage{amsmath}
\usepackage{amsfonts}
\usepackage{amssymb}
\usepackage{graphicx}
\usepackage{subfigure}

\begin{document}

\title{Deflection of Slow Light by Magneto-Optically Controlled Atomic Media}
\author{D.L. Zhou $^{1}$}
\email{zhoudl72@aphy.iphy.ac.cn}
\author{Lan Zhou$^{2}$}
\author{R.Q. Wang $^{1}$}
\author{S. Yi $^{2}$}
\author{C.P. Sun$^{2}$}
\email{suncp@itp.ac.cn}
\homepage{www.itp.ac.cn/~suncp}
\affiliation{$^{1}$Institute of Physics, Chinese Academy of Sciences, Beijing 100080,
China}
\affiliation{$^{2}$Institute of Theoretical Physics, Chinese Academy of Sciences, Beijing
100080, China}

\begin{abstract}
We present a semi-classical theory for light deflection by a
coherent $\Lambda$-type three-level atomic medium in an
inhomogeneous magnetic field or an inhomogeneous control laser. When
the atomic energy levels (or the Rabi coupling by the control laser)
are position-dependent due to the Zeeman effect by the inhomogeneous
magnetic field (or the inhomogeneity of the control field profile),
the spatial dependence of the refraction index of the atomic medium
will result in an observable deflection of slow signal light when
the electromagnetically induced transparency happens to avoid medium
absorption. Our theoretical approach based on Fermat's principle in
geometrical optics not only provides a consistent explanation for
the most recent experiment in a straightforward way, but also
predicts the new effects for the slow signal light deflection by the
atomic media in an inhomogeneous off-resonant control laser field.
\end{abstract}

\maketitle

\section{Introduction}

Many optical phenomena in nature, such as mirage and rainbow, can be
explained in terms of refraction of light rays in an inhomogeneous
optical medium \cite{BW99}. The theoretical approach can be
developed from Fermat's principle, saying that a light ray with a
given frequency traverses the path between two points which takes
the least time. Fermat's principle is consistent with light
traveling in a straight line in a medium with homogenous refraction
index. Classically, refraction of light results from the spatial
inhomogeneity of refraction index caused by the inhomogeneity of
medium density. In this article we show that the quantum coherence
of optical medium even with homogenous density can also result in a
spatially inhomogeneous refraction index and thus various phenomena
on refraction of light.

Light deflection by an atomic medium in external fields has been
studied experimentally in the last two decades
\cite{SW92,Hol97,PJAH04,Mos95}. A most recent experiment with a
rubidium atomic gas \cite{KW06} was carried out to demonstrate how
the electromagnetically induced transparency (EIT)
\cite{HFI90,BIH91} enhances the light deflection in a $\Lambda$-
type three-level atomic medium. Remarkably, if a magnetic field with
some gradient is applied to the gas cell, and the signal light and
the control light satisfies the frequency matching condition to
realize the EIT, the signal light will transmit perfectly through
the atomic media with a very slow group velocity, and thus be
deflected with an angle proportional to its propagation time through
the gas cell. A signal light can be deflected not only by atomic
media in a nonuniform magnetic field, but also by an atomic medium
coherently driven by an inhomogeneous laser field, which is shown in
another recent experiment \cite{SLRS07}. As an ultra-dispersive
optical prism, such coherence enhanced media has an angular
dispersion which is six orders of magnitude higher than that of a
prism made of optical glass.

The light deflection phenomenon observed in Ref. \cite{KW06} was
explained in terms of the dark polariton concept \cite{lukin,scp},
in which the quantized signal light field dresses the atomic
collective excitation to form a quasi-particle -- the polariton with
an effective magnetic moment. Then, the experimental result could be
understood as a Stern-Gerlach experiment for the quasi-particle. It
worths noticing that, in this explanation, the signal light must be
assumed as a quantized field.

\begin{figure}[h]
\includegraphics[width=8.5cm]{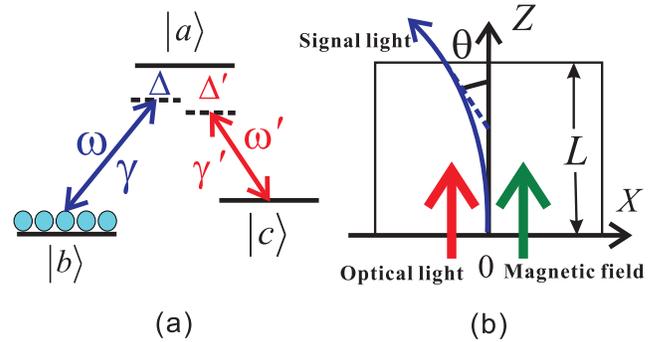}
\caption{\textit{(Color Online)} (a) Atomic level configuration: three
levels are coupled by a signal field and a control field with detunnings $%
\Delta$ and $\Delta^{\prime}$ respectively. (b) Schematic diagram of light
deflection in the atomic medium: a transverse magnetic field gradient causes
inhomogeneity of the refraction index of the medium via EIT and then
deflection of the signal light beam.}
\label{fig1}
\end{figure}

In this paper we present a semi-classical theory to uniformly treat
light deflection phenomena by a coherent $\Lambda$-type three-level
atomic medium in various external fields. Note that the EIT enhances
spatial dependence of the refraction index of the atomic medium,
which results from the inhomogeneity of the magnetic field or the
control laser field. We thus apply the Fermat's principle to
calculate the signal light path in such coherent medium where the
quantization of the signal light is not necessary to account for the
experiments.

Our theory is semi-classical since the atoms are described quantum
mechanically, while both the two laser fields are treated classically. This
approach not only provides a consistent explanation for the experiment \cite%
{KW06} in a straightforward way, but also predicts the new effects
for the weak light deflection by an atomic media driven by an
optical laser with inhomogeneous profile. This situation with weak
probe light is essentially different from the current experiment
about ``ultra-dispersive optical prism" \cite{SLRS07}, where the
strength of the probe light is as strong as that of the control
light, and thus the linear suspensibility obtained for weak probe
light will not be enough to account for the data of the experiment.

\section{Semi-classical approach for light deflection}

As shown in Fig. 1 (a),
the system under consideration is an atomic gas cell filled with $\Lambda $%
-type three-level atoms with an upper level $|a\rangle$ and two lower levels $%
|b\rangle$ and $|c\rangle$. The radiative decay rates from $|a\rangle$ to $%
|b\rangle$ and from $|a\rangle$ to $|c\rangle$ are, respectively, $\gamma$ and $%
\gamma^{\prime}$. The level splitting between $|a\rangle$ and
$|b\rangle$ (between $|a\rangle$ and $|c\rangle$) is denoted as
$\omega_{ab}$ ($\omega_{ac}$). We assume that $|a\rangle$ and
$|b\rangle$ are coupled by
a weaker `signal' field with the frequency $\omega$, while $|a\rangle$ and $%
|c\rangle$ are coupled by a stronger `control' field with the frequency $%
\omega^{\prime}$. The detunings of these two transitions are denoted
 as $\Delta$ and $\Delta^{\prime}$ respectively, which are defined by
$\Delta=\omega-\omega_{ab}$ and
$\Delta^{\prime}=\omega^{\prime}-\omega_{ac}$. The linear
susceptibility of the medium for the weak signal light can be
expressed as~\cite{SF92,SZ97,FIM05}
\begin{equation}
\chi=\chi_0\frac{\Gamma\delta\left(
|\Omega^{\prime}|^{2}-4\Delta\delta+i2\delta\Gamma\right) }
{(|\Omega^{\prime}|^{2}-4\Delta\delta)^{2}+4%
\delta^2\Gamma^2},   \label{chi1}
\end{equation}
where the constant $\chi_0=4N|d_{ab}|^2/(\epsilon_0\hbar\Gamma)$,
 $N$ is the density of atomic gas, $d_{ab}$ is the matrix element between
 the states $|a\rangle$ and $|b\rangle$ of the
 dipole operator,  $\Omega^{\prime}$ is the Rabi frequency of the control light,
$\Gamma=\gamma+\gamma^{\prime}$, and
 $\delta\equiv
\Delta-\Delta^{\prime}$ is the two-photon detuning.

In deriving Eq. (\ref{chi1}), the one-photon detuning $\Delta$, the
two-photon detuning $\delta$, and the Rabi coupling
$\Omega^{\prime}$ are assumed to be independent of spatial position.
When the external fields exerted on the atomic gas are
inhomogeneous, such as in the experiments \cite{KW06,SLRS07}, the
above parameters will become spatial dependent, which are denoted as
$\Delta(\vec{r})$, $\delta(\vec{r})$, and $\Omega^{\prime}(\vec{r})$
respectively.  Let us assume that the atomic gas cell can be divided
into many smaller cells, each smaller cell containing a large number
of atoms and the inhomogeneous external field being sufficiently
homogenous for each smaller cell. Thus we can apply Eq. (\ref{chi1})
to each smaller cell by changing the values of the parameters
$\Delta$, $\delta$, and $\Omega^{\prime}$ for different cells. When
the inhomogenous external fields are exerted on the atomic gas,
under the above approximation, the linear susceptibility in Eq.
(\ref{chi1}) will become spatial dependent:
\begin{equation}
\chi(\vec{r})=\chi_0\frac{\Gamma\delta(\vec{r})\left(
|\Omega^{\prime}(\vec{r})|^{2}-4\Delta(\vec{r})\delta(\vec{r})
+i2\delta(\vec{r})\Gamma\right) }
{\left(|\Omega^{\prime}(\vec{r})|^{2}-4\Delta(\vec{r})\delta(\vec{r})\right)^{2}+4%
{\delta(\vec{r})}^2{\Gamma}^2}.  \label{chi11}
\end{equation}

To grasp the main physics in Eq. (\ref{chi11}), we consider the case
where $\delta(\vec{r}), \Delta(\vec{r})\ll \Gamma \ll
|\Omega^{\prime}(\vec{r})|$. In the first order approximation, the
linear susceptibility is simplified to
\begin{equation}
\chi(\vec{r})=\chi_0\frac{\Gamma\delta(\vec{r})}
{|\Omega^{\prime}(\vec{r})|^{2}}. \label{chir}
\end{equation}
The vanishing of the imaginary part of the susceptibility in Eq.
(\ref{chir}) can be attributed to the steady dark state formed by
two lower atomic levels $|b\rangle$ and $|c\rangle$, which
completely eliminates the dissipation due
to the radiative decay of the excited state $|a\rangle$. Since $\chi(\vec {r}%
)\ll1$ near the two-photon resonance, the refraction index is
approximated as
\begin{equation}
n(\vec{r})\simeq1+\frac{1}{2}\chi(\vec{r}).   \label{aaa1}
\end{equation}

Once the refraction index $n(\vec{r})$ is known, the trajectory of a
light ray propagating in the atomic medium can be obtained by
solving the the differential equation \cite{BW99}
\begin{equation}
\frac{d}{ds}[n(\vec{r})\frac{d\vec{r}}{ds}]=\nabla n(\vec{r}),
\label{patheq}
\end{equation}
where $ds=\sqrt{dx^{2}+dy^{2}+dz^{2}}$. Inserting Eq. (\ref{aaa1})
into Eq. (\ref{patheq}), we get
\begin{equation}
\frac{d^2\vec{r}}{ds^2}+\left(\frac {\nabla\chi(\vec{r})} {2}
\cdot\frac{d\vec{r}}{ds}\right)\frac{d\vec{r}}{ds}=\frac
{\nabla\chi(\vec{r})} {2}. \label{patheq1}
\end{equation}
In the first order approximation, we can apply Eqs. (\ref{chir}) and
(\ref{patheq1}) to determine the light ray trajectory and the
corresponding deflection angle.

To demonstrate this procedure, let us consider an example related to
the experiments to be studied later. We assume that the signal light
injects at the position $\vec{r}_{i}=(x_{i},0,0)$ along $z$ axis. To
further simplify the calculation, we approximate the gradient of the
linear susceptibility along the light trajectory to be that at the
incident point, and the direction of this gradient along $x$ axis,
namely,
\begin{equation}
\frac{\triangledown \chi(\vec{r})} {2}
\simeq\frac{\triangledown\chi(\vec{r}%
_{i})}{2}=\frac {1} {\eta} \vec{e}_{x} \label{eta}
\end{equation}
with $\vec{e}_{x}$ being the unit vector along $x$ axis. Then Eq.
(\ref{patheq1}) allows an analytic solution of the light ray path
\begin{align*}
x(s) & =x(0)+\eta \ln \cosh \frac {s} {\eta}, \\
y(s) & =0, \\
z(s) &=\eta\sinh \frac {s} {\eta}.
\end{align*}
When the light ray exits the atomic gas cell, we have $z(s_{f})=L$,
where $L$ is the length along $z$ direction of the atomic gas cell.
The length of light path in the atomic medium is then given by
$s_{f}=\eta\sinh^{-1} (L/\eta)$, and we finally arrive at the light
deflection angle
\begin{equation}
\theta\equiv\frac{\dot{x}(s_{f})}{\dot{z}(s_{f})}=\frac
{L/\eta}{1+L^2/\eta^2 }\simeq \frac {L} {\eta}, \label{theta}
\end{equation}
where the last equality is satisfied only when $L\ll \eta$, which is
satisfied throughout this paper.

\section{Light propagation in optically controlled media}

Here we study the deflection of a weak signal light under a
spatially inhomogeneous control light $\Omega^{\prime}(\vec{r})$. At
first sight, this model may look similar to that considered in
experiment Ref.~\cite{SLRS07}. However, after carefully examing the
experimental parameters, we realized that the experiment was
performed using a stronger signal light, for which the linear
susceptibility theory is insufficient and the experimental results
cannot be explained using our semi-classical theory in the present
formulation.

\begin{figure}[ptb]
\includegraphics[width=5.5 cm]{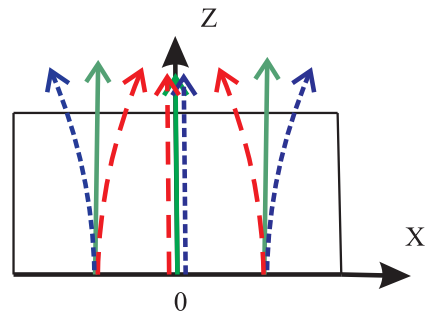}
\caption{\textit{(Color Online)} Schematic illustration about deflection of
week signal light in the presence of control light with inhomogeneous
profile as in Eq. ( \protect\ref{gauss}). The three cases with detuning $%
\protect\delta<0$, $\protect\delta=0$, and $\protect\delta>0 $ are
denoted by red, green, and blue color respectively. Corresponding to
these three cases the light rays with incidents in different
positions $x <0$, $x =0$, and $x >0 $ will possesses different
deflection ways. } \label{defang2}
\end{figure}

In the absence of an external magnetic field, the two-photon detuning $%
\delta$ is position independent. We further assume that the driving
light has a Gaussian profile
\begin{equation}
\Omega ^{\prime }(\vec{r})=\Omega _{0}^{\prime }\exp (-\frac{x^{2}+y^{2}}{%
\sigma ^{2}}),  \label{gauss}
\end{equation}%
with $\sigma $ characterizing the width of the profile. From Eq. (\ref{chir}%
), one can easily obtain the linear susceptibility for the signal light to
be
\begin{equation}
\chi (\vec{r})=\chi_0 \frac{\Gamma\delta}{|\Omega _{0}^{\prime
}|^{2}}\exp (2\frac{x^{2}+y^{2}}{\sigma ^{2}}). \label{chi2}
\end{equation}%
An immediate consequence of the above equation is that the light is
undeflected if the two photon detuning $\delta$ is set to zero. This
result clearly distinguishes our model from the experimentally studied case~%
\cite{SLRS07}, where a significant light deflection was observed
even at resonance. For $\delta \neq 0$ case, we consider the
situation where the signal light is sufficiently weak compared to
the control light and in the limit $\delta \ll \Omega ^{\prime
}(\vec{r})$. For the signal lights incident onto the medium at $\vec{r}%
_{i}=(x_{i},0,0)$ and along the positive $z$-axis within the region
$\sqrt{x^{2}+y^{2}}\lesssim \sigma$, we have
\begin{equation}
\frac{\triangledown\chi(\vec{r}%
_{i})}{2}=\chi_0 \frac{4 \Gamma\delta x_i}{|\Omega _{0}^{\prime
}|^{2}\sigma^2}\exp (\frac{2x_i^{2}}{\sigma ^{2}})\vec{e}_{x}=\frac
{1} {\eta} \vec{e}_{x}.
\end{equation}%
Eq. (\ref{theta}) immediately yields the deflection angle
\begin{equation}
\theta \simeq \chi_0 \frac{4 \Gamma\delta x_i L}{|\Omega
_{0}^{\prime }|^{2}\sigma^2}\exp (\frac{2x_i^{2}}{\sigma ^{2}}).
\label{theta3}
\end{equation}%
It follows from the above equation that, for $x_{i}\neq 0$, a red
detuned signal
light ($\delta <0$) feels an \textquotedblleft attractive potential" toward $%
z$ axis; while a blue detuned light ($\delta >0$) experiences a
\textquotedblleft repulsive potential". At $x_{i}=0$, the signal
light is undeflected irrespective of its detuning. We also note that
the deflection angle increase when $\Omega _{0}^{\prime }$ becomes
weaker. These novel light deflection phenomena are schematically
illustrated in Fig. \ref{defang2}.

\begin{figure}[h]
\includegraphics[width=8.5 cm]{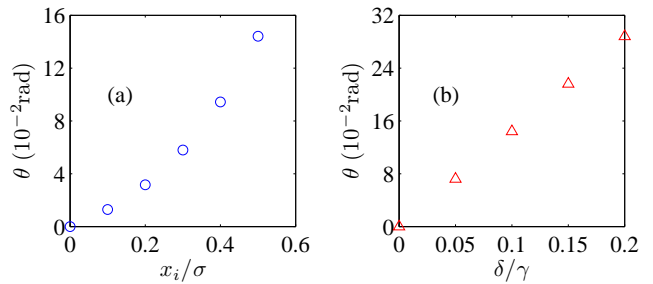}
\caption{\textit{(Color Online)} The deflection angle of the signal
light varies with (a) the injection position $x_{i}$
($\delta=0.1\Gamma$), (b) the two-photon detuning
($x_{i}=0.5\sigma$).} \label{defang4}
\end{figure}

To get a quantitative idea about the deflection angle, we calculate $\theta$
using the optimal experimental parameters given in Refs.\cite{KW06,SLRS07}.
For example, $\sigma=5mm$, $L=10\sigma$, $\Omega_{0}^{\prime}=5\Gamma$, and $%
N=10^{12}/cm^{3}$. As shown in Fig. \ref{defang4}, we see that the
deflection angle becomes larger as the inject position $x_{i}$
increases. Note that this result is valid only when the intensity of
the local control light $\Omega
_{0}^{\prime}\exp(-x_{i}^{2}/\sigma^{2})\gg\Gamma$. In addition, the
deflection angle increases linearly with the two-photon detuning
$\delta$. The deflection angle can reach $0.29\;\text{rad}$, which
is three orders larger than that in the previous experiment
\cite{KW06}. Thus, these interesting predictions are experimentally
observable, and can be explicitly tested by tuning the frequency of
the signal light and (or) the incidence position $x$ of the signal
light.

As being pointed out previously, the experiment results in Ref. \cite%
{SLRS07} cannot be directly explained using the semiclassical theory
based on susceptibility formula Eq. (\ref{chi2}). To fully account
for the experimental results, a different mechanism based on quantum
coherence should be considered. Since both the signal light and the
control field are strong, there should be the coherent population
trapping for the atoms with the EIT configuration \cite{SZ97}, which
will result in the nonlinear response for the EIT like effect. It is
believed that a refined theoretical approach accounting for the
strong signal light, and connecting the quantum interferences of the
atomic transitions is required.

\section{Magnetically controlled light deflection}

For the next example, we present a semiclassical explanation for the
light deflection by a
coherent atomic medium subjected to an inhomogeneous magnetic field \cite%
{KW06}, where the Rabi frequency $\Omega^{\prime}$
are uniform. We consider a linearized inhomogeneous magnetic field $\vec{B}%
(x)=(B_{0}+B_{1}x)\vec{e}_{z}$. Thus the two photon detuning%
\begin{equation}
\delta (\vec{r}) =\omega-\omega^{\prime}-\mu_{B} (B_{0}+B_{1}x),
\end{equation}
where $\mu_B$ is the Bohr magneton, and we used the related spectrum
data of ${}^{87}Rb$ $D1$ line \cite{Ste}. Thus Eq. (\ref{eta}) gives
\begin{equation}
\frac{\triangledown\chi(\vec{r}%
_{i})}{2}=-\chi_0 \frac{ \Gamma\mu_B B_1}{2|\Omega _{0}^{\prime
}|^{2}}\vec{e}_{x}=\frac {1} {\eta} \vec{e}_{x}.
\end{equation}
In the experiment~\cite{KW06}, the light injects onto the atomic gas
cell at the point $\vec{r}_{i}=(0,0,0)$ and along $z$ axis.
Following Eq. (\ref{theta}), we find the deflection angle to be
\begin{equation}
\theta=-\chi_0 \frac{ \Gamma\mu_B B_1 L}{2|\Omega _{0}^{\prime
}|^{2}}. \label{theta1}
\end{equation}
The deflection angle can be reexpressed in terms of group velocity. To this
end, we note that
\begin{equation}
v_{g}=\frac{c}{n+\omega dn/d\omega}\approx\frac{c}{\omega}
\frac{2|\Omega _{0}^{\prime }|^{2}}{\chi_0 \Gamma}, \label{vg}
\end{equation}
where we have utilized the fact that%
\begin{equation*}
\chi_0 \frac{ \Gamma\omega}{2|\Omega _{0}^{\prime }|^{2}}\gg 1.
\end{equation*}
Inserting (\ref{vg}) into Eq. (\ref{theta1}), we get the deflection
angle
\begin{equation}
\theta=-\frac{c\mu_{B}B_{1}L}{v_{g}\omega},   \label{theta2}
\end{equation}
which has exactly the same form as that obtained in Ref. \cite{KW06}
where the control light was treated quantum mechanically. Our
calculation, however, indicates that in weak field limit the
classical treatment on the control light is capable of capturing the
central result on light deflection in an EIT atomic gas.

Note that the concept of group velocity for a light ray  does not
play any role in our geometric optics method. The unique purpose to
derive Eq. (\ref{theta2}) is to compare Eq. (\ref{theta1}) with the
known result in Ref. \cite{KW06}. Although these two methods give
the same result on light deflection angle, the physical picture are
quite different. In the picture of dark polariton in Ref.
\cite{KW06,FL02}, the velocity of dark polariton is the group
velocity $c/(n+\omega dn/d\omega)$. In the picture of light ray,
however, the signal light with a given frequency propagates with the
phase velocity $c/n(\omega)$.

In addition, only transparent medium is considered in geometric
optics \cite{LLP84}. This is why we can correctly deal with the
signal light deflection in the EIT window, where the atomic gas is
transparent to the signal light. However, when we need to further
investigate the  phenomena on signal light deflection, (for example,
considering the signal light frequency region outside of the EIT
window, or the energy-conserving dephasing process of the atomic
level $|c\rangle$), the atomic medium becomes dissipative, and thus
the geometric optics method will not be valid any more. A better
solution is to directly solve the Maxwell equations in continuous
medium, which is beyond the scope of our article.

\section{Conclusion}

In summary, we have presented a semi-classical approach to describe
the light deflection in the atomic gas cell by applying an
inhomogeneous magnetic field or an inhomogeneous pump optical field.
Our theory not only explains the experiment without quantization of
the probe light, but also predicts some interesting phenomena on
quantum coherence enhanced light deflection. The EIT effect not only
makes the atomic medium transparent near the two-photon resonance,
but also makes the linear susceptibility of the atomic medium
spatial dependent. It is this spatial dependent linear
susceptibility that deflects the signal light ray. For applications
in quantum information processing, the EIT enhanced light deflection
can motivate a protocol for quantum sate storage with
spatially-distinguishable channels.

This work is supported by NSFC with grant Nos. 90203018, 10474104,
10674141, and 60433050, and NFRPC with Nos. 2001CB309310,
2005CB724508, 2006CB921206, and 2006AA06Z104. The authors
acknowledged helpful discussions with H. Wang, J.X. Zhang, and K.C.
Peng.

\end{document}